\documentclass[sigconf]{acmart}
\usepackage{enumitem}
\setlist[itemize]{leftmargin=*}
\usepackage{multirow}
\usepackage{makecell}
\AtBeginDocument{%
  \providecommand\BibTeX{{%
    \normalfont B\kern-0.5em{\scshape i\kern-0.25em b}\kern-0.8em\TeX}}}

\setcopyright{acmcopyright}
\copyrightyear{2018}
\acmYear{2022}

\acmConference[]{}{}{}
\acmPrice{15.00}
\acmISBN{978-1-4503-XXXX-X/18/06}

\begin{document}

\title{Knowledge graph enhanced recommender system}



\author{${\rm Zepeng\ Huai^{1,2}}$, ${\rm Jianhua\ Tao^{1,2,3}}$, ${\rm Feihu\ Che^{1,2}}$, ${\rm Guohua\ Yang^{1}}$, ${\rm Dawei\ Zhang^{1}}$}
\affiliation{%
  \institution{${\rm ^{1}National\ Laboratory\ of\ Pattern\ Recognition, Institute\ of\ Automation,\ Chinese\ Academy\ of\ Sciences}$}
  \institution{${\rm ^{2}School\ of\ Artificial\ Intelligence,\ University\ of\ Chinese\ Academy\ of\ Sciences}$}
  \institution{${\rm ^{3}CAS\ Center\ for\ Excellence\ in\ Brain\ Science\ and\ Intelligence\ Technology}$}
  \city{Beijing}
  \country{China}
  \postcode{43017-6221}
}
\email{huaizepeng2020@ia.ac.cn,jhtao@nlpr.ia.ac.cn, chefeihu2017@ia.ac.cn, dawei.zhang@nlpr.ia.ac.cn, guohua.yang@nlpr.ia.ac.cn}

%
%

\renewcommand{\shortauthors}{}
\begin{abstract}
  (Since KDD requires the papers that have been submitted to arXiv at least one month prior to the deadline need a different title and abstract, here we give a brief version of title and abstract.) Knowledge Graphs (KGs) have shown great success in recommendation. This is attributed to the rich attribute information contained in KG to improve item and user representations as side information. However, existing knowledge-aware methods leverage attribute information at a coarse-grained level both in item and user side. In this paper, we proposed a novel \textit{attentive knowledge graph attribute network}(AKGAN) to learn item attributes and user interests via attribute information in KG. Technically, AKGAN adopts a heterogeneous graph neural network framework, which has a different design between the first layer and the latter layer. With one attribute placed in the corresponding range of element-wise positions, AKGAN employs a novel interest-aware attention network, which releases the limitation that the sum of attention weight is 1, to model the complexity and personality of user interests towards attributes. Experimental results on three benchmark datasets show the effectiveness and explainability of AKGAN.
\end{abstract}




\maketitle

\section{Introduction}
Recommender systems have shown great success in e-commerce, online advertisement, and social medial platforms. The core task of recommender systems is to solve the overload information problem~\cite{wang2018billion,ying2018graph,wang2019knowledge} and suggest items that users are potentially interested in. Traditional methods to achieve information filtering are content-based and collaborative filtering (CF)-based recommender systems\cite{ma2008sorec,he2017neural,wang2019neural}, which utilize items’ content features and the similarity of users or items from interaction data respectively\cite{guo2020survey}. However, both of them don't introduce much side information.

\begin{figure}[!tb]
	\centering
	\includegraphics[width=\linewidth]{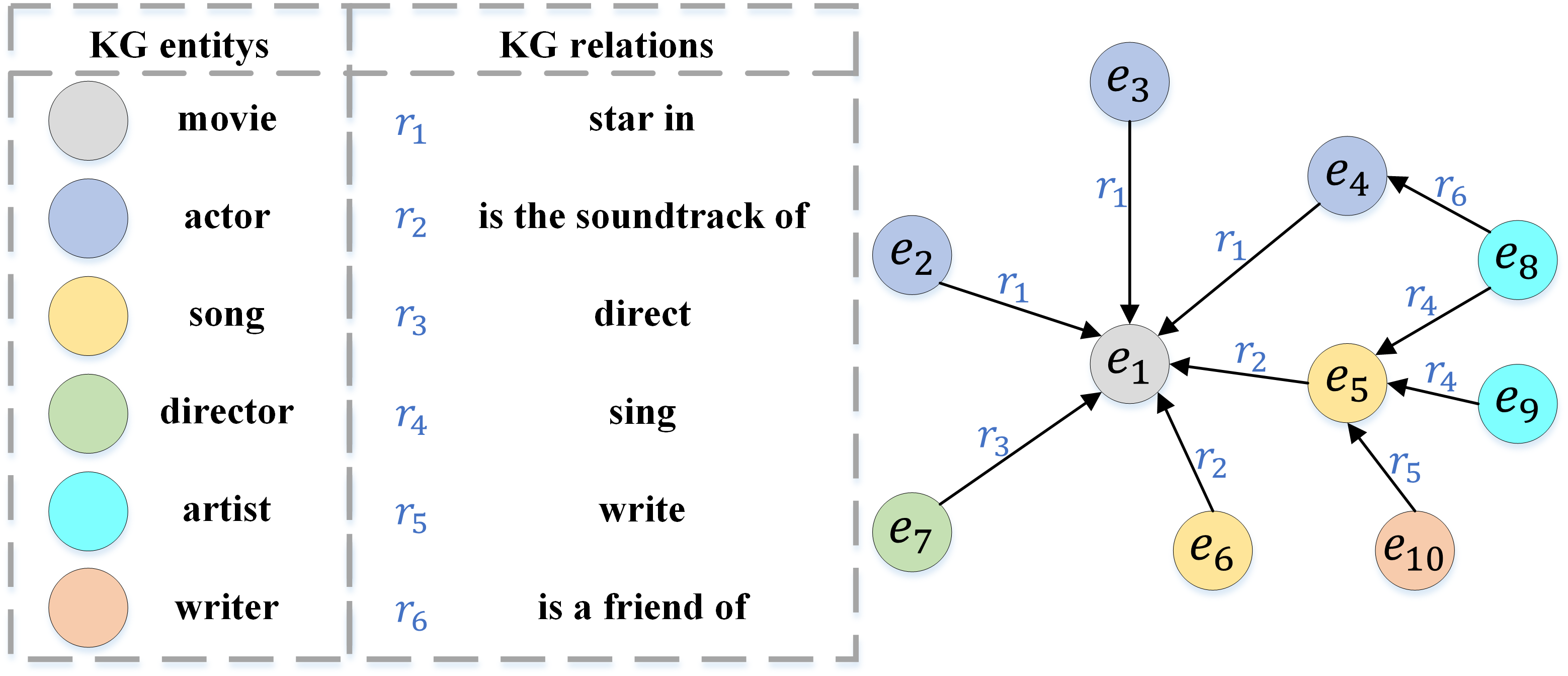}
	\caption{An example of how KG contains multiple attributes information. Best viewed in color.}
	\Description{A woman and a girl in white dresses sit in an open car.}
	\label{fig:introduction figure1}
\end{figure}

In recent years, introducing knowledge graphs (KGs) into recommender systems as side information has been effective for improving recommendation performance. KGs represent real-world entities and illustrate the relationship between them with graph data structure, which can reveal multiple attributes of items and explore the potential reason for user-item interactions. Generally, a KG contains item nodes and attribute nodes, and attribute nodes can not only describe items' attributes directly but also represent other attribute nodes' attributes. For example, a movie knowledge graph (as shown in figure \ref{fig:introduction figure1}) has six types of nodes, where movie node (i.e., $e_1$) is item node and the others (i.e., $e_{2-10}$) are attribute nodes. $e_{2-4}$ represent the actor attribute of movie $e_1$, which means actors $e_{2-4}$ stared in movie $e_1$. And $e_{8,9}$ represent the singer attribute of song $e_5$, which means singer $e_{8,9}$ sang the song $e_5$. To integrate attribute information into recommender system, earlier works focus on embedding-based methods \cite{ai2018learning,cao2019unifying,zhang2016collaborative,huang2018improving,zhang2018learning} and path-based methods \cite{yu2013collaborative,yu2013recommendation,yu2014personalized,luo2014hete,shi2015semantic}. The former exploits KGs with knowledge graph embedding (KGE) algorithm (i.e., TransE \cite{bordes2013translating} and
TransH \cite{wang2016text}) to learn entity embeddings and then feed them into a recommender framework. The latter usually predefines a path scheme (i.e., meta-path) and leverages path-level semantic similarities of entities to refine the representations of users and items. However, both of them don't capture high-order connectivities and fail to exploit both the rich semantics and topology of KGs. More recently, the propagation-based methods, a.k.a. graph neural network (GNN)-based methods such as KGAT\cite{wang2019kgat}, KGNN-LS\cite{wang2019knowledge1}, KNI\cite{qu2019end}, AKGE\cite{sha2019attentive}, KGIN\cite{wang2021learning}, have attracted considerable interest of researchers. With the attribute information iteratively propagating in KGs, GNN-based methods integrate multi-hop neighbors into representations and have achieved the state-of-the-art recommendation results.

Despite the success of GNN in exploiting multi-hop attribute information, we argue that there are still three shortcomings: (1)\textbf{The pollution caused by weighted sum operation in merging different attribute information}. Different attributes are independent in terms of semantics and user preference. Take the actor attribute $e_2$ and the singer attribute $e_8$ of movie $e_1$ in figure \ref{fig:introduction figure1} as an example, semantics independence means $e_2$ doesn't co-occur with $e_8$ in each movie, since an actor and a singer are invited to work for a movie respectively. Preference independence means whether a user prefers actor $e_2$ is independent with whether he likes singer $e_8$. However, existing GNN-based methods pool embeddings from different attribute nodes with weighted sum (i.e., sum, mean, attention) operation, which brings about semantic pollution and the difficulty to distill user interested attribute information. (2) \textbf{The nonlinearity between the distance and importance of attribute node relative to item node}. Usually, high-order neighbors are less relative to the center node in graph data, but this is not absolute. For example, singer attribute $e_8$ is more valued than director attribute $e_7$ by some movie viewers who like music, while $e_8$, a 2-hop neighbor, is further than $e_7$, a 1-hop neighbor. This problem is caused by inherent graph topology, which means some significant attribute nodes don't connect to item nodes directly, such that it requires multiple passes to integrate these high-order neighbors to the center node. However, existing GNN-based algorithms neglect this issue and decrease the weight of significant high-order neighbors coupled with the increase of propagation times. (3)\textbf{The complexity and personality of user interests towards different attributes}. User interests show the following pattern: a user just gets interested in a part of rather than all attributes of an item, and different users prefer different attributes even towards the same item. For example, both user $u_1$ and $u_2$ watch movie $e_1$ because $u_1$ likes $e_1$'s actor $e_2$ rather than director $e_7$ while $u_2$ prefers the theme song $e_5$ rather than actor $e_2$. Therefore, we should integrate actor $e_2$ rather than director $e_7$ attribute embedding into $u_1$'s representation and integrate song $e_5$ rather than actor $e_2$ attribute embedding into $u_2$'s representation. In other words, the item representation learned by GNN propagation contains noisy signals and we should distill part attributes interested by user personally. However, existing GNN-based algorithms recognize this pattern insufficiently, like \cite{wang2019kgat} doesn't consider noisy attributes and \cite{wang2021learning} doesn't consider personal extraction.

To address the foregoing problems, we propose a novel \textit{attentive knowledge graph attribute network} (AKGAN), which consists of two components: (1) \textbf{knowledge graph attribute network (KGAN)}. The core task of KGAN is to learn informative item representations without semantic pollution and weight decrease of significant attribute nodes. Technologically, KGAN has a different design between the first layer and the latter layer under GNN framework. The first layer, called attribute modeling layer, aims to generate initial item representations without semantic pollution. It regards each relation in KG as an attribute and has two key designs: embedding each entity in different attribute spaces and using concatenation operation to merge different attribute information. The latter layer, called attribute propagation layer, aims to remain semantics unpolluted and avoids weight decrease of significant neighbors after GNN propagation. Finally, KGAN generates item representations where one attribute is represented within a specific range of element-wise positions independently. (2) \textbf{user interest-aware attention network}. With different attributes placed in corresponding element-wise positions, we design an interest-aware attention layer to distill user interested attributes. Specifically, for a particular user, we pool the representations of his interacted items and extract each attribute representation according to corresponding element-wise positions. Then each attribute representation will be fed into an attention module to calculate a personal interest score, which describes how much he prefers this attribute. We introduce a novel activation unit to release the limitation that the sum of attention weight is 1, which aims to reserve the intensity of user interests\cite{zhou2018deep}. Finally all interest scores and item representations are further combined to infer user representations. 

To summarize, the main contributions of this work are as follows:
\begin{itemize}
	\item On the item side, we propose a novel knowledge graph attribute network, which employs a heterogeneous design in different layers under GNN framework, embeds each entity in different attribute spaces, and combines different attributes via concatenation operation, to avoid pollution caused by weighted sum operation and weight decrease of significant neighbors.
	\item On the user side, we develop an interest-aware attention network, which introduces a novel activation unit and releases the limitation that the sum of attention weight is 1 , to model user interests towards attributes personally.
	\item We conduct extensive experiments on three public benchmarks, and the results demonstrate the superior performance of AKGAN over state-of-the-art baselines. In-depth analyses are provided to illustrate the interpretability of AKGAN for user personal interests.
\end{itemize}

%
%

\section{Problem Formalizaiton}
We begin by introducing some related notations and then defining the KG enhanced recommendation problem. Let $\mathcal{U}=\{u\}$ and $\mathcal{I}=\{i\}$ separately denote the user and item sets. A typical recommender system usually has historical user-item interactions, which is defined as $\mathcal{O}^{+}=\{(u,i) | u \in \mathcal{U} ,i \in \mathcal{I} \}$. Each $(u,i)$ pair indicates user $u$ has interacted with item $i$ before, such as clicking, review, or purchasing. We also have a knowledge graph which stores the structured semantic information of real-world facts. We denote the entity and relation sets in KG as $\mathcal{R}=\{r\}$ and $\mathcal{E}=\{e\}$. KG is presented as $\mathcal{G}=\{(h,r,t)|h\in \mathcal{E},r \in \mathcal{R},t\in \mathcal{E}\}\}$, where each tripile means there is a relation $r$ from head entity $h$ to tail entity $t$. For example, $(James\ Cameron, Direct, Avader)$ describes the fact that James Cameron is the director of the movie Avatar. Note that $\mathcal{R}$ contains relations in both canonical direction (e.g., Direct) and inverse direction (e.g., DirectedBy). The bridge of KG and recommender system is that items are contained in entities. Specificlly, $\mathcal{E}$ consists of item node $\mathcal{I}$($\mathcal{I} \subseteq \mathcal{E}$) as well as attribute node $\mathcal{E} \setminus \mathcal{I}$, which can futher refine user representation $\mathbf{e}_u$ and item representation $\mathbf{e}_i$. 

We now formulate the KG enhanced recommendation task to be addressed in this paper:
\begin{itemize}
	\item \textbf{Input}: a knowledge graph $\mathcal{G}$, which contains rich sturcture senmeantic information of item, and the user-item interaction data $\mathcal{O}^{+}$
	\item \textbf{Output}: a scoring function that denotes the probability that user u will interact with item $i$.
\end{itemize}


\section{METHODOLOGY}
\subsection{Model Overview}
\begin{figure*}[h]
	\centering
	\includegraphics[width=\linewidth]{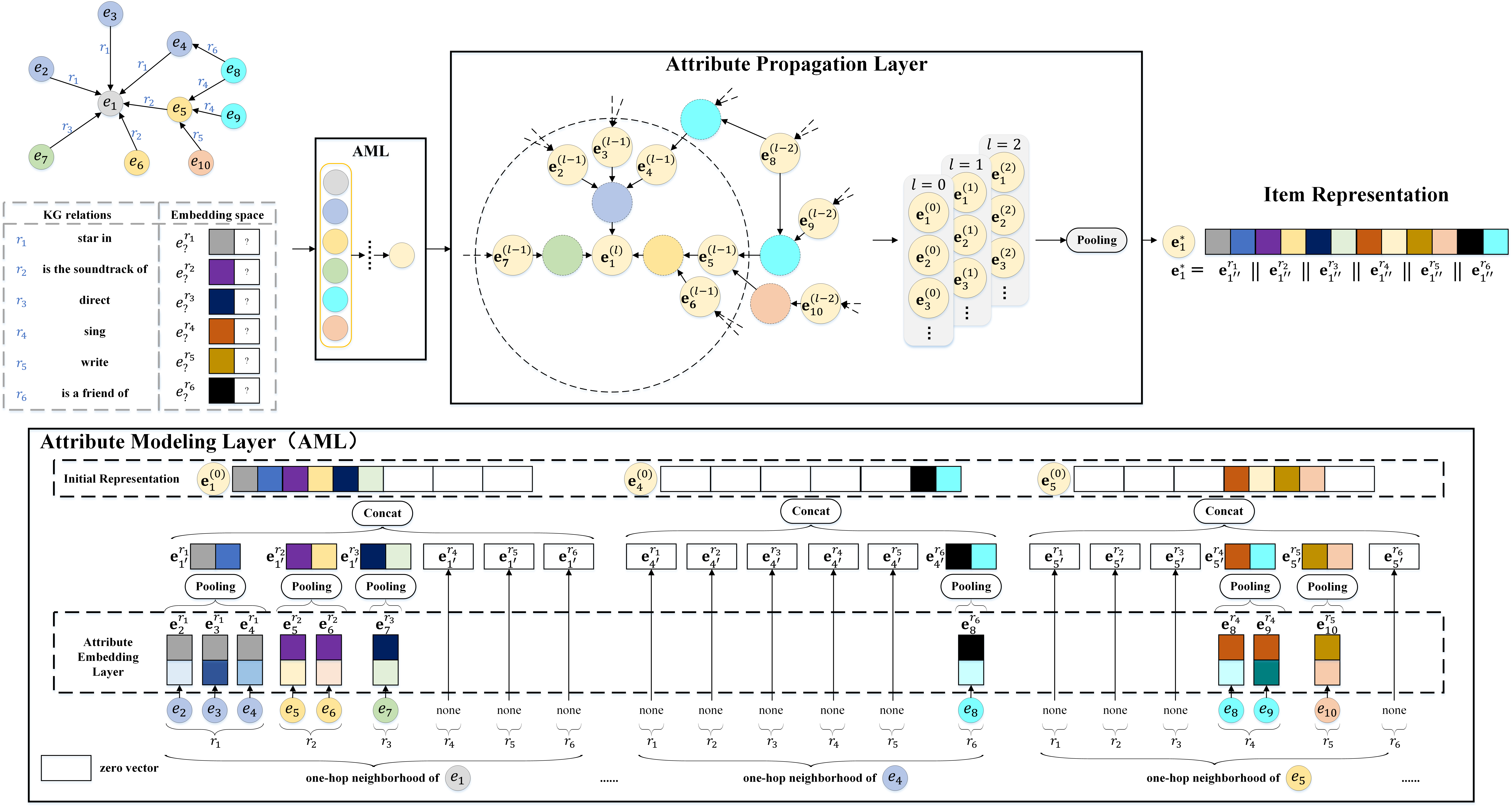}
	\caption{The structure of KGAN.}
	\Description{A woman and a girl in white dresses sit in an open car.}
	\label{fig:KGAN}
\end{figure*}

\begin{figure*}[h]
	\centering
	\includegraphics[width=0.6\linewidth]{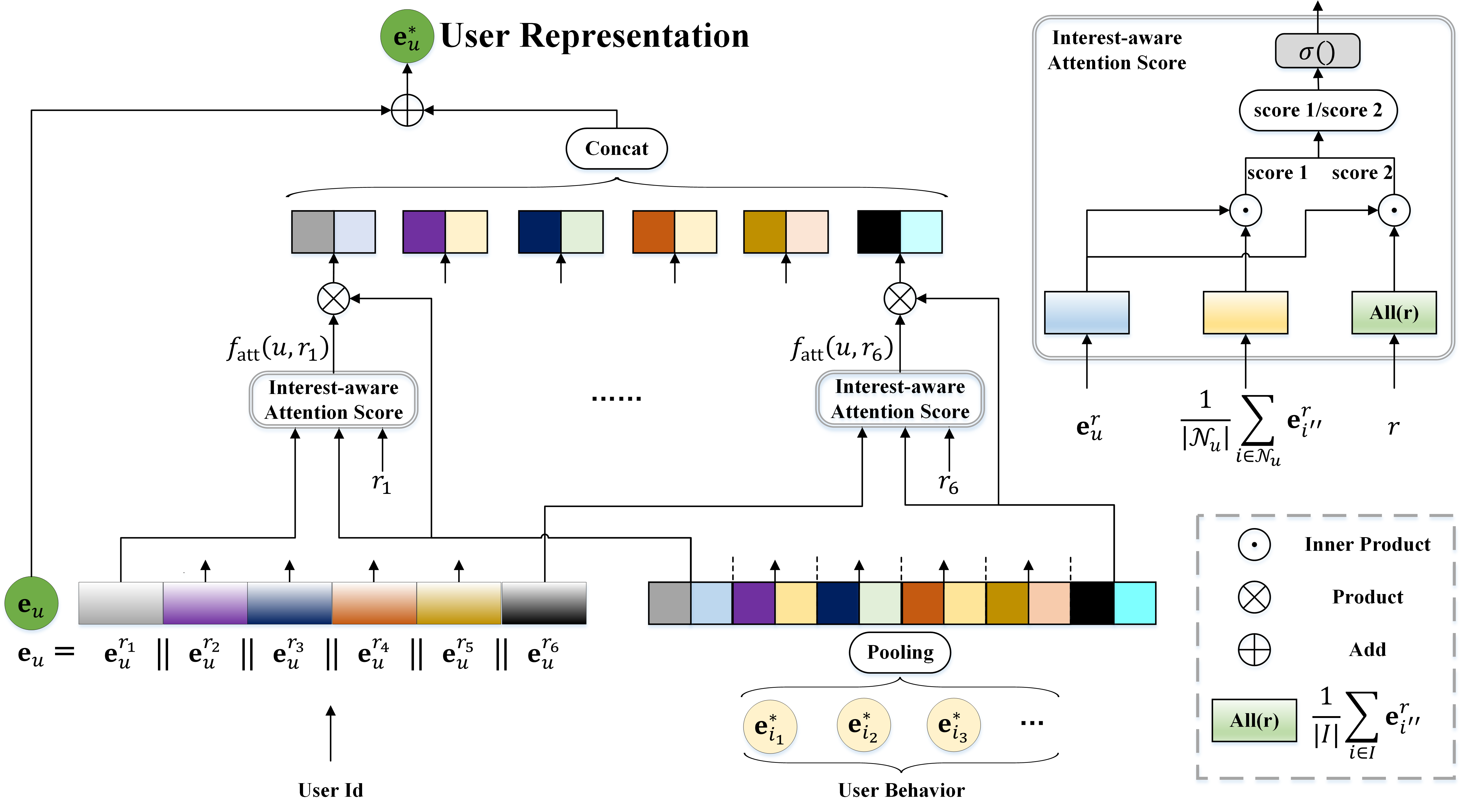}
	\caption{The structure of IAAN.}
	\Description{A woman and a girl in white dresses sit in an open car.}
	\label{fig:IAAN}
\end{figure*}

In this subsection, we introduce the framework of AKGAN. With a widely used two-tower structure in recommender model\cite{zhang2021model,niu2020dual,he2020lightgcn}, AKGAN consists of two main modules: (1) knowledge graph attribute network, which is illustrated in Figure \ref{fig:KGAN}, and (2) user interest-aware attention module, which is illustrated in Figure \ref{fig:IAAN}. 

The core task of KGAN is to learn KG enhanced item representations that contain multiple attribute information. KGAN adopts GNN framework and has a heterogeneous design in the first layer and the latter layer, which are named attribute modeling layer and attribute propagation layer respectively.

\textbf{Attribute modeling layer} (AML) is to construct initial entity representations by merging one-hop neiborhoods' attribute information in knowledge graph, as follows:
\begin{equation}
	\label{eq:KGAN0_1}
	\begin{aligned}
		\mathbf{e}^{(0)}_i=f_{{\rm AML}}(\mathbf{e}^{atr}_{i},\mathcal{G})
	\end{aligned}
\end{equation}

\textbf{Attribute propagation layer} (APL) recursively propagates attribute information to acquire more informative item representations as
\begin{equation}
	\label{eq:KGAN0_2}
	\begin{aligned}
		\mathbf{e}^{(l)}_i=f_{{\rm APL}}(\mathbf{e}^{(l-1)}_i,\mathcal{G})
	\end{aligned}
\end{equation}

After performing $L$ layers, we obtain multiple item representations, namely $\{\mathbf{e}^{(0)}_{i},...,\mathbf{e}^{(L)}_{i} \}$. As the output of $l_{{\rm th}}$ layer represents $l-$hop attribute information, we conduct sum opration to pool them and infer final item representations as 
\begin{equation}
	\label{eq:KGAN0_f}
	\begin{aligned}
		\mathbf{e}^{*}_i=\sum_{l=0}^{L}\mathbf{e}^{(l)}_{i}
	\end{aligned}
\end{equation}

When item representations have been learned, \textbf{interest-aware attention} (IAA) module is to learn user representations via interaction data with a novel relation-aware attention mechanism which represents user interests towards relations, a.k.a. attributes. Formally, user representations are obtained by IAA as
\begin{equation}
	\label{eq:IAA0}
	\begin{aligned}
		\mathbf{e}^{*}_u=f_{{\rm IAA}}(\mathbf{e}_i,\mathcal{O}^{+})
	\end{aligned}
\end{equation}

When user representations and item representations are learned, a scoring function is used to predict their matching score and here we adopt inner product operation as
\begin{equation}
	\label{eq:scoref0}
	\begin{aligned}
		\hat{y}(u,i)={\mathbf{e}^{*}_u}^\top\mathbf{e}^{*}_i
	\end{aligned}
\end{equation}

%
%
%
%

\subsection{Attribute Modeling Layer}
In KGs, one entity has different types of relations with neighbors. For example, in figure \ref{fig:KGAN}, $e_8$ has different relations with $e_4$ and $e_5$. To model a pure initial representation without semantic pollution, we regard each relation as an attribute and embeds each entity in all attribute embedding spaces, which is similar to the design of FFM\cite{juan2016field} that embeds each feature in different fields. Therefore each entity has sevecal embeddings and we denote embedding set in different attribute spaces as
\begin{equation}
	\label{eq:embed_atr}
	\begin{aligned}
		\mathcal{E}^{atr}=\{(\mathbf{e}^{r_1}_i,\mathbf{e}^{r_2}_i,...,\mathbf{e}^{r_M}_i)|i \in \mathcal{E},M=\lvert \mathcal{R} \rvert \}
	\end{aligned}
\end{equation}
where $\mathbf{e}^{r_m}_i \in \mathbb{R}^{d^{m}}$ denotes the embedding of entity $i$ in attribute $r_m$ space and $d^m$ is the embedding dimension of attribute $r_m$ space.

Then we construct initial entity representations by aggregating attribute embeddings of one-hop neighbors. We can see that each neighbor has several embeddings and just one relation-aware embedding will be used. Taking figure \ref{fig:KGAN} as an example, there are two links $(e_8,r_4,e_5)$ and $(e_8,r_6,e_4)$, we use embeddings $\mathbf{e}^{r_4}_8$ to represents the singer attribute of song $e_5$ and embedding $\mathbf{e}^{r_6}_8$ to represents the friend attribute of actor $e_4$, respectively. After we prepared the relation-aware embeddings, we adopt average operation to pool the same relation-aware neighbors to acquire the main semantics of this attribute as
\begin{equation}
	\label{eq:atr_represtt_1}
	\begin{aligned}
		\mathbf{e}^{r_j}_{i^{\prime}}=\frac{1}{\mid \mathcal{N}_i^{r_j} \mid} \sum_{j \in \mathcal{N}_i^{r_j} } \mathbf{e}^{r_j}_j
	\end{aligned}
\end{equation}
where $j \in \mathcal{N}_i^{r_j}$ and $\mathcal{N}_i^{r_j}=\{j|j \in (j,r_j,i),(j,r_j,i)\in \mathcal{G}\}$ denotes the set of head entities that belong to the triplet where $i$ is tail entity and $r_j$ is relation. For example, if $e_{2,3}$ are both comedy actors and $e_4$ is an action actor, such that $mean(\mathbf{e}^{r_1}_{2-4})$ represents movie $e_1$ is likely to show much funny performance. Note that not all attributes occur in one-hop range, like movie $e_1$ doesn't has friend attribute $r_6$. To obtain a fixed-length representation, we provide zero vector to the absent attribute. Finally, all attribute embeddings will be integrated into one representation by concatenation operation as
\begin{equation}
	\label{eq:atr_represtt_2}
	\begin{aligned}
		\mathbf{e}^{(0)}_i=\mathbf{e}^{r_1}_{i^{\prime}}\left\vert\kern-0.25ex\right\vert\mathbf{e}^{r_2}_{i^{\prime}}\left\vert\kern-0.25ex\right\vert\cdots\left\vert\kern-0.25ex\right\vert\mathbf{e}^{r_M}_{i^{\prime}}
	\end{aligned}
\end{equation}

Note that we adopt concatenation rather than weighted sum operation (i.e., sum, mean, attention), which aims to solve the problem of semantic pollution. More specfically, equation \ref{eq:atr_represtt_2} shows that one attribute is represented within a specific range of element-wise positions. In addition, one type of attribute is placed in the same element-wise postions for all entities. For example, two representations $\mathbf{e}^{(0)}_1=\mathbf{e}^{r_1}_{1^{\prime}}\left\vert\kern-0.25ex\right\vert\cdots\left\vert\kern-0.25ex\right\vert\mathbf{e}^{r_3}_{7}\left\vert\kern-0.25ex\right\vert\cdots$ and $\mathbf{e}^{(0)}_{11}=\mathbf{e}^{r_1}_{11^{\prime}}\left\vert\kern-0.25ex\right\vert\cdots\left\vert\kern-0.25ex\right\vert\mathbf{e}^{r_3}_{7}\left\vert\kern-0.25ex\right\vert\cdots$ mean movie $e_1$ and movie $e_{11}$ have the same director and different actors. In a word, a concatented representation avoids the interaction and preserves semantic independence of different attributes, which will be further advantageous to maintain the weight of siginificant high-order neighbors in subsection \ref{subsec:APL} and distill user interested attributes in subsection \ref{subsec:IAA}.

%
%
%
%

\subsection{Attribute Propagation Layer}\label{subsec:APL}
When we obtain initial entity representations via one-hop neighbors, an intuitive idea is to propagate them via GNN framework, so as to aggregate high-order neighbors into the center node and acquire more informative item representations.

Here we regard a KG as a heterogeneous graph and adopt a widely used two-step scheme\cite{wang2019heterogeneous,zhang2019heterogeneous,fu2020magnn} to aggregate the representations of neighbors: (1) same relation-aware neighbors aggregation; (2) relations combination.

The same relation-aware neighbors contains the similiar attribute types. For example, in figure \ref{fig:KGAN}, both $e_2$ and $e_3$ are actors and contain friend attribute while $e_5$ is a song and contains singer attribute. Therefore, we firstly aggregate representations of the same relation-aware neighbors. Secondly, to learn a more comprehensive representation, we need to fuse multiple attributes hold in different relation-aware neighbors. The pooling methods in the above two steps are average and sum operation, respectively. And we leave the further exploration of other pooling methods like attention as the future work. More formally, in the $l-$th layer, we recursively formulate the representation of an entity as:
\begin{equation}
	\label{eq:atr_pro_1}
	\begin{aligned}
		\mathbf{e}^{(l)}_i=\sum_{r_j \in \mathcal{R}} \frac{1}{\mid \mathcal{N}_i^{r_j} \mid} \sum_{j \in \mathcal{N}_i^{r_j} } \mathbf{e}^{(l-1)}_j
	\end{aligned}
\end{equation}

Now the final item representation $\mathbf{e}_i$ has been learned by equation \ref{eq:KGAN0_f}, let's check how KGAN avoids pollution and weight decrease of significant high-order neighbors. 

We re-examine $\mathbf{e}^{*}_i$ from the perspective of element-wise position. Without loss of generality,  $\mathbf{e}_i$ is contructed as
\begin{equation}
	\label{eq:atr_f_0}
	\begin{aligned}
		\mathbf{e}^{*}_i=\mathbf{e}^{r_1}_{i^{\prime\prime}}\left\vert\kern-0.25ex\right\vert\mathbf{e}^{r_2}_{i^{\prime\prime}}\left\vert\kern-0.25ex\right\vert\cdots\left\vert\kern-0.25ex\right\vert\mathbf{e}^{r_M}_{i^{\prime\prime}}
	\end{aligned}
\end{equation}
where $\mathbf{e}^{r_m}_{i^{\prime\prime}} \in \mathbb{R}^{d^{m}}$ is a truncated vector in $\mathbf{e}^{*}_i \in \mathbb{R}^{D}$ and $D=\sum\limits_{r_m \in \mathcal{R}}d^{m}$. The start and ending index of $\mathbf{e}^{r_m}_{i^{\prime\prime}}$ in $\mathbf{e}^{*}_i$ are
$\sum\limits_{p=1}^{m-1}d^p$ and $\sum\limits_{p=1}^{m}d^p$, respectively. Reviewing the generation process of $\mathbf{e}^{r_m}_{i^{\prime\prime}}$, we can see that  $\mathbf{e}^{r_m}_{i^{\prime\prime}}$ is learned by $i$'s neighbors' embeddings in attribute $r_m$ space and doesn't contain any embeddings from other attribute spaces. For example, in figure \ref{fig:KGAN}, $\mathbf{e}^{r_1}_{1^{\prime\prime}}$ is learned by $\mathbf{e}^{r_1}_{2}$, $\mathbf{e}^{r_1}_{3}$, and $\mathbf{e}^{r_1}_{4}$. This means $\mathbf{e}^{r_m}_{i^{\prime\prime}}$ maintains the independence of attribute $r_m$ and KGAN remains semantics unpolluted after multi-layer propagations. Furthermore, we check sepecific neighbors which are aggregated into $\mathbf{e}^{r_m}_{i^{\prime\prime}}$. Take $\mathbf{e}^{r_5}_{1^{\prime\prime}}$ as an exapmle, after 1-order propagation (one AML), $\mathbf{e}^{r_5}_{1^{\prime\prime}}$ is a zero vector since attribute $r_5$ doesn't occur in one-hop neighbors of movie $e_1$. Then after 2-order propagation (one AML and one APL), $\mathbf{e}^{r_5}_{1^{\prime\prime}}=\mathbf{e}^{r_5}_{10}$, which seems as if there were a link $(e_{10},r_5,e_1)$ and $e_1$ connected to $e_{10}$ directly. In general, when one relation, a.k.a attribute (i.e., $r_5$), firstly appear in the receptive field of center node (i.e., $e_1$) at $l-$hop (i.e., $2-$hop) position, the weight of its corresponding node (i.e., $e_{10}$) will not been decreased, which proves that KGAN maintains the weight of significant high-order neighbors.

%

\subsection{Interest-aware Attention Layer}\label{subsec:IAA}
After item representations have been obtained by KGAN, a typical idea in recommender system is to enhance user representations by clicked items, like \cite{koren2008factorization,wu2019neural,wang2018dkn}. We take avearage pooling as an example as
\begin{equation}
	\label{eq:IAA_1}
	\begin{aligned}
		\mathbf{e}^{*}_u=\mathbf{e}_{u}+\frac{1}{\mid \mathcal{N}_u \mid}\sum_{i \in \mathcal{N}_u}\mathbf{e}^{*}_i
	\end{aligned}
\end{equation}
where $\mathcal{N}_u=\{i|(u,i)\in \mathcal{O}\}$ and $\mathbf{e}_{u}$ represents user embedding for collaborative filtering. However, avearage pooling doesn't consider user preferences personally, thereby recent works aims to model user interests via attention mechanism, such as \cite{wang2019kgat,li2019multi,qin2020user,wang2021learning,zhou2019deep}.

Here we propose a novel attention module to model user interests towards different attributes. We assign an interest score $f_{{\rm att}}(u,r_m)$ to each pair of attribute $r_m$ and user $u$, and personally generate user representation by combining interest scores and interacted items. We firstly introduce how to calculate $f_{{\rm att}}(u,r_m)$ and then illustrate how to combine $f_{{\rm att}}(u,r_m)$ with interacted items. 

With different attributes placed in corresponding element-wise positions as shown in equation \ref{eq:atr_f_0}, we truncate user vector $\mathbf{e}_{u}$ with the same strategy as
\begin{equation}
	\label{eq:IAA_3}
	\begin{aligned}
		\mathbf{e}_{u}=\mathbf{e}^{r_1}_{u}\left\vert\kern-0.25ex\right\vert\mathbf{e}^{r_2}_{u}\left\vert\kern-0.25ex\right\vert\cdots\left\vert\kern-0.25ex\right\vert\mathbf{e}^{r_M}_{u}
	\end{aligned}
\end{equation}
where $\mathbf{e}^{r_m}_{u} \in \mathbb{R}^{d^{m}}$ is a truncated vector in $\mathbf{e}_{u} \in \mathbb{R}^{D}$. The start and ending index of $\mathbf{e}^{r_m}_{u}$ in $\mathbf{e}_{u}$ are
$\sum\limits_{p=1}^{m-1}d^p$ and $\sum\limits_{p=1}^{m}d^p$, respectively. Then the interest score of user $u$ towards attribute $r_m$ is calculated by
\begin{equation}
	\label{eq:IAA_4}
	\begin{aligned}
		f_{{\rm att}}(u,r_m)=\sigma \left( \tau \frac{\frac{1}{\mid \mathcal{N}_u \mid} \sum_{i \in \mathcal{N}_u}{\mathbf{e}^{r_m}_{u}}^\top \mathbf{e}^{r_m}_{i^{\prime\prime}}}{\frac{1}{\mid \mathcal{I} \mid} \sum_{i \in \mathcal{I}}{\mathbf{e}^{r_m}_{u}}^\top \mathbf{e}^{r_m}_{i^{\prime\prime}}} \right)
	\end{aligned}
\end{equation}
where $\tau$ is temperature coefficient as a hyperparameter. In equation \ref{eq:IAA_4}, $\frac{1}{\mid \mathcal{N}_u \mid} \sum_{i \in \mathcal{N}_u}{\mathbf{e}^{r_m}_{u}}^\top \mathbf{e}^{r_m}_{i^{\prime\prime}}$ denotes the interest-degree of user $u$ for his interacted items in attribute $r_m$, while $\frac{1}{\mid \mathcal{I} \mid} \sum_{i \in \mathcal{N}_u}{\mathbf{e}^{r_m}_{u}}^\top \mathbf{e}^{r_m}_{i^{\prime\prime}}$ denotes the interest-degree of user $u$ for all items in attribute $r_m$. If user $u$ attachs great importance to attribute $r_m$ when choosing items, the corresponding truncated representation $\mathbf{e}^{r_m}_{i^{\prime\prime}}$ of his interacted items will be radically different from that of other uninterested items, such that  $\frac{1}{\mid \mathcal{N}_u \mid} \sum_{i \in \mathcal{N}_u}{\mathbf{e}^{r_m}_{u}}^\top \mathbf{e}^{r_m}_{i^{\prime\prime}}$ will be much greater than $\frac{1}{\mid \mathcal{I} \mid} \sum_{i \in \mathcal{N}_u}{\mathbf{e}^{r_m}_{u}}^\top \mathbf{e}^{r_m}_{i^{\prime\prime}}$, and vice versa. Therefore, the ratio between the above two expressions denotes the interest-degree of user $u$ for attribute $r_m$. Since a negative value of ratio is meaningless, we first feed this ratio to ${\rm Relu}$ and then select $\tanh$ as the nonlinear activation function, therefore $\sigma(x)=\tanh( {\rm Relu} (x))$.

After $f_{{\rm att}}(u,r_m)$ has been prepared, we need to combine it with interacted items to learn user representations. We firstly re-express equation \ref{eq:IAA_1} as
\begin{equation}
	\label{eq:IAA_5}
	\begin{aligned}
		\mathbf{e}^{*}_u=\mathbf{e}_{u}+\mathop{\left\vert\kern-0.25ex\right\vert}\limits_{r_m \in \mathcal{R}} \left(\frac{1}{\mid \mathcal{N}_u \mid} \sum_{i \in \mathcal{N}_u}\mathbf{e}^{r_m}_{i^{\prime\prime}} \right)
	\end{aligned}
\end{equation}
where $\left\vert\kern-0.25ex\right\vert$ is the concatenation operation. Therefore an intuitive idea is to use $f_{{\rm att}}(u,r_m)$ to control how much attribute information, a.k.a. $\frac{1}{\mid \mathcal{N}_u \mid}\sum_{i \in \mathcal{N}_u}\mathbf{e}^{r_m}_{i^{\prime\prime}}$, will be passed to user. Consider the limit
case, when $f_{{\rm att}}(u,r_m)=0$ which means user $u$ doesn't pay attention to attribute $r_m$ at all, any information of attribute $r_m$ should not be contained in user representaion. Formally, user representation is obtained by
\begin{equation}
	\label{eq:IAA_6}
	\begin{aligned}
		\mathbf{e}^{*}_u=\mathbf{e}_{u}+\mathop{\left\vert\kern-0.25ex\right\vert}\limits_{r_m \in \mathcal{R}} \left( \frac{f_{{\rm att}}(u,r_m)}{\mid \mathcal{N}_u \mid} \sum_{i \in \mathcal{N}_u}\mathbf{e}^{r_1}_{i^{\prime\prime}} \right)
	\end{aligned}
\end{equation}

Note that here we relax the constraint that the sum of attention weights towards all attributes is 1, a.k.a. $\sum_{r_m \in \mathcal{R}}{f_{{\rm att}}(u,r_m)} \neq 1$. The reason is as follows: when the number of members participating in the attention calculation is large and the limitation that the sum of attention weights is 1 is still reserved at this time, it will cause the attention weight of each member to be dispersed, making it difficult to learn the coefficients of important nodes. This problem has also appeared in Graphair \cite{hu2021graphair}, which shows estimating $\mathcal{O}({\mid \mathcal{N} \mid}^2)$ coefficients exposes the risk
of overfitting. Therefore, we introduce the above novel activation unit to release the limitation that the sum of attention weight is 1, and the same idea is adopted by DIN \cite{zhou2018deep}.
%

\subsection{Model Optimization}
With the user representation $\mathbf{e}^{*}_u$ and the item representation $\mathbf{e}^{*}_i$ ready, equation \ref{eq:scoref0} is adopted to calculate the prediction score of each pair of user and item. Then we employ the BPR loss \cite{rendle2012bpr} to encourage that the observed interactions should be assigned higher prediction values than unobserved ones. The
objective function is formulated as
\begin{equation}
	\label{eq:loss}
	\begin{aligned}
		\mathcal{L}=\sum_{(u,i^{+},i^{-})\in \mathcal{O}} -\ln \sigma \bigg( \hat{y}(u,i^{+})-\hat{y}(u,i^{-}) \bigg) +\lambda\|\Theta\|_{2}^{2}
	\end{aligned}
\end{equation}
where $\mathcal{O}=\{ (u,i^{+},i^{-}) | (u,i^{+}) \in \mathcal{O}^{+}, (u,i^{-}) \in \mathcal{O}^{-}\}$ denotes the training set, which contains the observed interactions $\mathcal{O}^{+}$ and the unobserved interactions $\mathcal{O}^{-}$; $\sigma(\cdot)$ is the sigmoid function. $\Theta=\{ \mathbf{e}_i^{r_m},\mathbf{e}_{u}|i \in \mathcal{I},u \in \mathcal{U}, r_m \in \mathcal{R}\}$ is the model
parameter set. $L_2$ regularization parameterized by $\lambda$ on $\Theta$ is conducted to prevent overfitting. We employ the
Adam \cite{kingma2014adam} optimizer and use it in a mini-batch manner.

\section{expirement}
We evaluate our proposed AKGAN method on three benchmark datasets to answer the following research questions: 
\begin{itemize}
	\item \textbf{RQ1}: Does our proposed AKGAN outperform the state-of-the-art
	recommendation methods?
	\item \textbf{RQ2}: How do different components (i.e., attribute modeling layer, attribute propagation layer, and interest-aware attention layer)
	affect AKGAN?
	\item \textbf{RQ3}: Can AKGAN provide potential explanations about user preferences towards attributes?
\end{itemize}

\subsection{Experimental Settings}
\hspace*{\fill} \ 

\noindent\textbf{Dataset Description}. We choose three benchmark datasets to evaluate our method: Amazon-Book\footnote{http://jmcauley.ucsd.edu/data/amazon}, Last-FM\footnote{https://grouplens.org/datasets/hetrec-2011/}, and Alibaba-iFashion\footnote{https://drive.google.com/drive/folders/1xFdx5xuNXHGsUVG2VIohFTXf9S7G5veq}. The former two datasets are released in \cite{wang2019kgat} and the last one is released in \cite{wang2021learning}, and all of them are publicly available. Each dataset consists of two parts: user-item interactions and a corresponding knowledge graph. The basic statistics of the three datasets are presented in Table \ref{tab:Statistics of the datasets}. We follow the same data partition used in \cite{wang2019kgat,wang2020reinforced} to split the datasets into training and testing sets. For each observed user-item interaction, we randomly sample one negative item that the user has not interacted with before, and pair it with the user as a negative instance.
\begin{table}[]
	\caption{Statistics of the datasets. 'Int./user' indicates the average number of interactions per user.}
	\label{tab:Statistics of the datasets}
	\begin{tabular}{c|ccc}
		\hline
		\multicolumn{1}{c|}{Datasets}  & Amazon-book & Last-FM & Alibaba-iFashion \\
		\hline\hline
		\multicolumn{1}{c|}{\# users} & 70,679 & 23,566 & 114,737\\
		\multicolumn{1}{c|}{\# items} & 24,915 & 48,123 & 30,040 \\
		\multicolumn{1}{c|}{\# interactions} & 847,733 & 3,034,796 & 1,781,093 \\
		\multicolumn{1}{c|}{\# Int./user} & 11.99 & 128.78 & 15.52 \\
		\hline\hline
		\multicolumn{1}{c|}{\# entities} & 88,572 & 58,266 & 59,156 \\
		\multicolumn{1}{c|}{\# relations} & 39 & 9 & 51 \\
		\multicolumn{1}{c|}{\# triples} & 2,557,746 & 464,567 & 279,155\\
		\hline           
	\end{tabular}
\end{table}

\hspace*{\fill} \ 

\noindent\textbf{Evaluation Metrics}. We evaluate our method in the task of top-K recommendation. For each user, we treat all the items that the user has not interacted with as negative and the observed items in the testing set as positive. Then we rank all these items and adopt two widely-used evaluation protocols: Recall@$K$ and NDCG@$K$, where $K$ is set as 20 by default. We report the average metrics for all users in the testing set.

\hspace*{\fill} \ 
\begin{table*}[t]
	\caption{Overall Performance Comparison. The best performance is boldfaced; the runner up is labeled with '*'. '$\%$Imp.' indicates the improvements. }
	\label{tab:Overall Performance Comparison}
	\begin{tabular}{|c|c|c|c|c|c|c|c|c|c|}
		\hline 
		\multirow{2}{*}{\text{Dataset}} & \multirow{2}{*}{\text{Metrics}} & KG-free&embedding-based&path-based&\multicolumn{4}{c|}{GNN-based}&\multirow{2}{*}{\text{Imp.}}\\
		\cline{3-9}
		& & MF & CKE & RippleNet & KGAT & KGNN-LS & KGIN & AKGAN & \\  
		\hline 
		\multirow{2}{*}{Amazon-Book}&\multirow{1}{*}{Recall}
		&0.1241&0.1287&0.1355&0.1473&0.1389& $0.1687^{*}$& \textbf{0.1783}&5.69\%\\
		\cline{2-10}
		&\multirow{1}{*}{NDCG}
		&0.0650&0.0674&0.0763&0.0782&0.0614&$0.0915^{*}$&\textbf{0.0994}&8.63\%\\
		\hline 
		\multirow{2}{*}{Last-FM}&\multirow{1}{*}{Recall}
		&0.0774&0.0780&0.0842&0.0876&0.0877&$0.0978^{*}$&\textbf{0.1209}&23.62\%\\
		\cline{2-10}
		&\multirow{1}{*}{NDCG}
		&0.0669&0.0659&0.0766&0.0745&0.0653&$0.0848^{*}$&\textbf{0.1066}&25.71\%\\
		\hline 
		\multirow{2}{*}{Alibaba-iFashion}&\multirow{1}{*}{Recall}
		&0.0921&0.1068&0.1121&0.1015&0.1046&$0.1147^{*}$&\textbf{0.1253}&9.24\%\\
		\cline{2-10}
		&\multirow{1}{*}{NDCG}
		&0.0562&0.0633&0.0695&0.0616&0.0582&$0.0716^{*}$&\textbf{0.0801}&11.87\%\\
		\hline 
	\end{tabular}
\end{table*}


\noindent\textbf{Baselines}. To demonstrate the effectiveness, we compare AKGAN with KG-free (MF), embedding-based (CKE), path-based (RippleNet), and GNN-based (KGAT, KGNN-LS, KGIN) methods:
\begin{itemize}
	\item \textbf{MF} \cite{rendle2012bpr}: This is matrix factorization optimized by the Bayesian personalized ranking (BPR) loss, which only considers the user-item	interactions and leaves KG untouched.
	\item \textbf{CKE} \cite{zhang2016collaborative}: This method uses TransR \cite{lin2015learning}, a typical knowledge graph embedding algorithm, to regularize the representations of items, which are fed into MF framework for recommendation.
	\item \textbf{RippleNet} \cite{wang2018ripplenet}: This model combines embedding-based methods and path-based methods to propagate users’
	preferences on the KG for recommendation. RippleNet first assigns entities in the KG with	initial embeddings using TransE and represents a user via entities related to his historically clicking items.
	\item \textbf{KGAT} \cite{wang2019kgat}: This method encodes user behaviors and item knowledge as an unified knowledge graph to exploit high-order connectivity. KGAT applies an attentive neighborhood aggregation mechanism on a holistic graph and introduces TransR to regularize the representations.
	\item \textbf{KGNN-LS} \cite{wang2019knowledge}: It uses a user-specific relation scoring function to transform a heterogeneous KG into a user-personalized weighted graph and employs label smoothness regularization to avoid overfitting of edge weights.
	\item \textbf{KGIN} \cite{wang2021learning}: KGIN is the state-of-the-art GNN-based recommender. It models each intent as an attentive combination of KG relations to explore intents behind user-item interactions and adopts a novel relational path-aware aggregation scheme.
\end{itemize}

\hspace*{\fill} \ 

\noindent\textbf{Parameter Settings}. We implement our AGKAN model in Pytorch and Deep Graph Library (DGL)\footnote{https://github.com/dmlc/dgl}, which is a Python package for deep learning on graphs. We released all implementations (code, datasets, parameter settings, and training logs) to facilitate reproducibility. The embedding size of one attribute space in AGKAN varies from 4 to 64, and the detailed design will be introduced in Appendix \ref{subsec:embedding dimension}. For a fair comparison, we fix the size of ID embeddings as 64, which equals the max embedding size of AKGAN, for all baselines, except RippleNet 32 due to its high computational cost. We adopt Adam \cite{kingma2014adam} as the optimizer and the batch size is fixed at 1024 for all methods. We use the Xavier initializer \cite{glorot2010understanding} to initialize model parameters. We apply a grid search for hyper-parameters: the learning rate is searched in $\{10^{-4}, 10^{-3}, 10^{-2}, 10^{-1}\}$, the coefficient of L2 normalization is tuned in $\{ 10^{-5}, 10^{-4}, \cdots , 10^{-1}\}$, the number of GNN layers $L$ is searched in $\{1, 2, 3\}$ for GNN-based methods, and the dropout ratio is tuned in $\{$0.0, 0.1, · · · , 0.9$\}$. Besides, we use the node dropout technique for KGAT, KGIN, and AKGAN, where the ratio is searched in $\{$0.0, 0.1, · · · , 0.9$\}$. For RippleNet, we set the number of hops as 2, and the memory size as 5, 15, 8 for Alibaba-iFashion, Last-FM, and Amazon-book respectively to obtain the best performance. Since RippleNet is a model for CTR prediction, we generate top-K items with top-K scores in all items, which are compared with the test set to compute  Recall@$K$ and NDCG@$K$. For KGAT, we use the pre-trained ID embeddings of MF as the initialization, which is also adopted by KGIN. Moreover, early stopping strategy is performed, i.e., premature stopping if Recall@20 on the test set does not increase for 10 successive epochs. 

\subsection{Performance Comparison(RQ1)} \label{subsec: RQ1}
We report the empirical results in Table \ref{tab:Overall Performance Comparison} where
we highlight the results of the best baselines (starred) and our
AKGAN (boldfaced).  And we also use $\%$Imp. to denote the percentage of relative improvement on each metric. The observations are as followed:


\begin{itemize}
	\item AKGAN consistently achieves the best performance on three datasets in terms of all measures. Specifically, it achieves significant	improvements over the strongest baselines w.r.t. NDCG@20 by 8.63\%, 25.71\%, and 11.87\% in Amazon-Book, Last-FM, and Alibaba-iFashion, respectively. These improvements are attributed to the following reasons: (1) By combining all attributes using concatenation operation, AKGAN avoids semantic pollution caused by weighted sum operation and learns more high-quality item representations for recommendation. (2) Compared to GNN-based baselines (i.e., KGAT, KGNN-LS, KGIN), AKGAN maintains the weight of significant high-order neighbors by placing different attributes in corresponding element-wise positions. (3) Benefiting from our novel interest-aware attention module that assigns an interest score to each pair of user and attribute, AKGAN can recognize the pattern of user interest at a fine-grained level to conduct a better personal recommendation.
	\item Jointly analyzing AKGAN across the three datasets, we find that the
	improvement on Last-FM is more significant than that on	Alibaba-iFashion and Amazon-Book. The main reason is that the interaction number per user of
	Last-FM (128.78) is much larger than that of the other two datasets (11.99, 15.52). Therefore, there exists richer interaction information on the Last-FM dataset for AKGAN to refine user and item representation by collaborative signals. This indicates that AKGAN will fully realize its potential in recommendation scenarios with dense interaction data.
	\item KG-free method (MF) underperforms knowledge-aware methods (i.e., CKE, RippleNet, AKGAN). A clear reason is MF doesn't leverage the rich attribute information in KG. 
	\item GNN-based methods (i.e., KGAT, KGNN-LS, KGIN, AKGAN) achieve better performance than path-based (RippleNet) and embedding-based (CKE) methods. A possible reason is that these three kinds of recommenders adopt different usages of attributes. GNN-based methods aggregate neighbors' attribute information into item nodes to learn more informative representations. The other two methods have a common limitation that both of them don't break loose from employing knowledge graph embedding algorithms to model attribute information and regularize node representations. 
	\item In four GNN-based methods, AKGAN performs best, KGIN is the second-best, while KGAT and KGNN-LS are at the same level and achieve the worst results. The decreasing performance is because that the level of how a recommender learns item attributes and user interests is in descending order, from fine-grained to coarse-grained. AKGAN uses concatenation operation to avoid attribute interaction while other models adopt weighted sum (attentive combination) operation. AKGAN  maintains the weight of significant high-order neighbors while other models neglect this issue. To achieve the above two advantages, AKGAN adopts a heterogeneous design in different layers while the others use a homogeneous GNN framework. AKGAN, KGIN, and KGNN-LS all learn user interests towards attributes explicitly while KGAT doesn't. The above comparison is listed in Table \ref{tab:GNN-based models Comparison}.
\end{itemize}
\begin{table}[t]
	\caption{GNN-based models Comparison: (1) CM - Combination methods of different attributes (Con - concatenation, WS - weighted sum); (2) WD - whether to avoid weight decrease of significant high-order neighbors; (3) GF - the GNN framework that a recommender adopts (Het - heterogeneous design in different layers, Hom - identical network in each layer; (4) IA - whether to learn user interests towards attributes; (5) level - the level of how a recommender learns item attributes and user interests (FG - fine-grained, CG - coarse-grained).}
	\label{tab:GNN-based models Comparison}
	\begin{tabular}{|c|c|c|c|c|c|c|c|}
		\hline 
		\multirow{2}{*}{\text{Model}} & \multicolumn{2}{c|}{CM} &\multirow{2}{*}{WD}&\multicolumn{2}{c|}{GF}& \multirow{2}{*}{IA}& \multirow{2}{*}{\text{level}}\\
		\cline{2-3}
		\cline{5-6}
		& Con & WS & &Het&Hom & & \\  
		\hline
		AKGAN& $\surd$ & $\times$ &$\surd$ &$\surd$& $\times$& $\surd$ &  \multirow{4}{*}{\makecell[c]{FG \\ $\downarrow$ \\ CG}}\\
		\cline{1-7} 
		KGIN& $\times$ & $\surd$ & $\times$ & $\times$&$\surd$ & $\surd$ &\\
		\cline{1-7}  
		KGNN-LS& $\times$ & $\surd$ & $\times$& $\times$&$\surd$ & $\surd$ &\\
		\cline{1-7} 
		KGAT& $\times$ & $\surd$ & $\times$& $\times$&$\surd$ & $\times$ &\\
		\hline 
	\end{tabular}
\end{table}
\subsection{Study of AKGAN(RQ2)}
In this section, we first conduct an ablation study to investigate the effect of cancatenation operation and interest-aware attention layer. Towards the further analysis, we study the influence of layer numbers. In what follows, we explore how the hyperparameter, a.k.a. temperature coefficient, affects the performance. 

\hspace*{\fill} \ 

\noindent\textbf{Impact of cancatenation operation \& interest score}. To demonstrate the necessity of cancatenation oepration and interest score, we compare the performance of AKGAN with the following three variants: (1) combing different attributes with average operation and discarding interest score, termed AKGAN-mean, (2) combing different attributes with sum operation and discarding interest score, termed AKGAN-sum, (3) only discarding interest score, termed AKGAN-noatt. Discarding interest score means we use equation \ref{eq:IAA_1} to learn user representations. Note that we don't adopt weighted sum operation and interest-aware attention layer simultaneously. The reason is that cancatenation operation is the prerequisite of interest-aware attention layer, therefore we have to discard interest score when testing weighted sum operation. The results are shown in Table \ref{tab:Impact of cancatenation operation and interest score} and we summarize the major findings as below:
\begin{table}[t]
	\caption{Impact of cancatenation operation and interest score.}
	\label{tab:Impact of cancatenation operation and interest score}
	\small
		\begin{tabular}{l|cc|cc|cc}
		\hline & \multicolumn{2}{c|}{Amazon-Book} & \multicolumn{2}{c|}{Last-FM } & \multicolumn{2}{c}{Alibaba-iFashion} \\
		& Recall & NDCG & Recall & NDCG & Recall & NDCG \\
		\hline \hline 
		AKGAN-mean & 0.1504 & 0.0799 & 0.0849 & 0.0713 & 0.1064 & 0.0660 \\
		AKGAN-sum  & 0.1489 & 0.0784 & 0.0886 & 0.0736 & 0.1045 & 0.0643 \\
		AKGAN-noatt  & 0.1768 & 0.0969 & 0.1141 & 0.1012 & 0.1173 & 0.0739 \\
		\hline
	\end{tabular}
\end{table}
\begin{itemize}
	\item Comparing AKGAN-mean/AKGAN-sum with AKGAN-noatt, we can clearly see that replacing concatenation operation with weighted sum operation dramatically degrades performance of recommendation, which indicates the superiority of cancatenation operation.
	\item The improvement from AKGAN-noatt to AKGAN verifies the necessity of interest score.
	\item Jointly comparing AKGAN-noatt and AKGAN across the three datasets, we find that the improvement on Last-FM is more significant than that on	Alibaba-iFashion and Amazon-Book, which is consistent with the aforesaid conclusion in subsection \ref{subsec: RQ1}. The main reason is that Last-FM has more dense interaction data than the other two datasets, such that AKGAN can better learn user interest in Last-FM.
\end{itemize}
\hspace*{\fill} \ 

\noindent\textbf{Impact of model depth}. We investigate the influence of depth of receptive field in AKGAN by searching $L$ in the range of $\{1, 2, 3\}$. Particularly, $L=1$ means AKGAN has only one attribute modeling layer. The results are reported in Table \ref{tab:Impact of the number of layers}. In Amazon-Book and Last-FM, we can see that increasing propagation times of attributes can boost the performance, because more attribute information is aggregated into the center node to learn more informative user and item representations. While in Alibaba-iFashion, AKGAN-2 is the best and AKGAN-3 is the worst, which is caused by its inherent topology. Alibaba-iFashion just has two kinds of triplets: the first-order
connectivity of an item is its components, a.k.a. \textit{(fashion outfit, including, fashion staff)}, and the second-order connectivity is the category of staff, a.k.a. \textit{(staff, having-category-?, ?)}. Therefore, all significant attribute information has been captured in the 2-hop range, leading to the best performance of AKGAN-2.

\begin{table}[t]
	\caption{Impact of the number of layers $L$. }
	\label{tab:Impact of the number of layers}
	\small
	\begin{tabular}{l|cc|cc|cc}
		\hline & \multicolumn{2}{c|}{Amazon-Book} & \multicolumn{2}{c|}{Last-FM } & \multicolumn{2}{c}{Alibaba-iFashion} \\
		& Recall & NDCG & Recall & NDCG & Recall & NDCG \\
		\hline \hline 
		AKGAN-1  & 0.1737 & 0.0974 & 0.1192 & 0.1060 & 0.1245 & 0.0791 \\
		AKGAN-2  & 0.1752 & 0.0977 & 0.1205 & 0.1069 & 0.1253 & 0.0801 \\
		AKGAN-3  & 0.1783 & 0.0994 & 0.1209 & 0.1066 & 0.1221 & 0.0776 \\
		\hline
	\end{tabular}
\end{table}

\hspace*{\fill} \ 

\noindent\textbf{Impact of temperature coefficient}.
We vary the temperature coefficient in the range
of $\tau=\{0.01,0.1,0.2,...,0.9,1\}$ to study its influence on the performance of AKGAN. The experiment is ongoing.  

\begin{figure*}[h]
	\centering
	\includegraphics[width=0.8\linewidth]{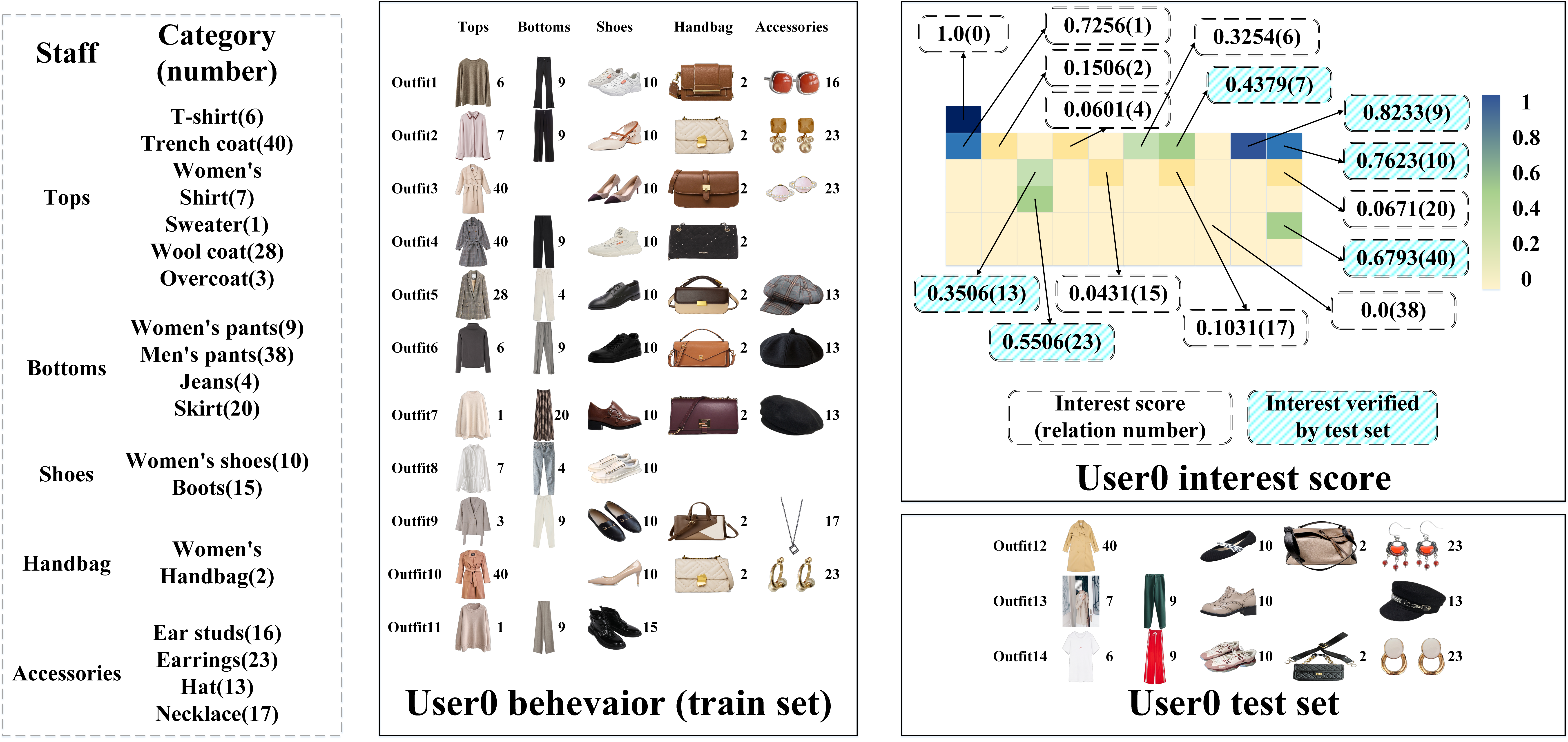}
	\caption{Case study.}
	\Description{visualize the interest score to show how AKGAN learn user preference towards attribute.}
	\label{fig:case}
\end{figure*}
\subsection{Case Study(RQ3)}
In this section, we visualize the interest score to show how AKGAN learns user preferences towards attributes. We choose Alibaba-iFashion as the example dataset since it helps to provide an intuitive explanation. We begin with introducing this dataset briefly for further better understanding. Alibaba-iFashion is an E-commerce dataset, which contains user-outfit click history for recommendation. Each outfit consists of several fashion staffs (e.g., tops, bottoms, shoes) and each staff is assigned with different categories. For example, trench coat, T-shirt and sweater are three kinds of tops, pants and long skirt are two kinds of bottoms. Alibaba-iFashion regards these categories as relations in KG and there is one more relation between outfit and staff, called \textit{including}. Therefore, Alibaba-iFashion KG has two kinds of triples: \textit{(outfit, including, staff)} and \textit{(staff, having-category-?, ?)}, where ? is the specific category like T-shirt. According to Table \ref{tab:Statistics of the datasets}, Alibaba-iFashion KG has 51 relations, where we number relation \textit{including} as 0 and number 50 staff categories from 1 to 50. In figure \ref{fig:case}, we first exhibit the training set and testing set of a specific user (user0). Then we visualize the interest score learned from the training set by AKGAN and use the testing set to validate the effectiveness of these scores. From figure \ref{fig:case} we can see that:
\begin{itemize}
	\item The interest score accurately models the user preferences towards attributes. Among 11 outfits clicked by user0, the tops contain four main kinds: T-shirt, shirt, trench coat, and sweater, which means user0 prefers the above four types of tops. The corresponding interest score are 0.3254, 0.4379, 0.6793, 0.7256, respectively. The most clicked bottoms are women's pants which are assigned with a high score 0.8223, while the jeans and skirt occur less frequently and their scores, 0.0601 and 0.0671, are particularly small. As for shoes, user0 prefers women's shoes and boots only appear once, their scores, 0.7623 and 0.0431, also reflect this difference. The earrings and hat are two main accessories and both of them are assigned with high scores, 0.5506 and 0.3506.
	\item All outfits clicked by user0 have only one kind of handbag and the interest score $f_{{\rm att}}(u_0,r_2)=0.1506$, which is a small value.  A possible reason is that outfits are manually created by Taobao’s fashion experts who prefer to include handbags for the completion of the outfit \cite{chen2019pog}. Therefore user0 judges whether to click an outfit based on other staffs rather than the handbag. Another explanation is that handbag is not as frequently changed as other staffs like tops in daily dressing, such that user0 pays less attention to the handbag when browsing through outfits.
	\item Testing set proves the effectiveness of interest score. For example, $f_{{\rm att}}(u_0,r_{40})=0.6793$ is proved to be reansonable since outfit12 which contains a trench coat appears in the testing set. And $f_{{\rm att}}(u_0,r_{9})=0.8233$ is validated by outfit13 and outfit14 with same reason.
	\item In one staff, the interest degree varies according to category. Take accessories as an example, user0 prefers earrings than hat and $f_{{\rm att}}(u_0,r_{23})=0.5506>f_{{\rm att}}(u_0,r_{13})=0.3506$. This result is also proved by the testing set, where two outfits contain earrings while only one outfit has a hat.
	\item Unconcerned attribute is assigned with a zero score, like $f_{{\rm att}}(u_0,r_{38})=0$. The reason is that user0 perhaps is female since her click histories are all women's clothing and she isn't interested in men's pants.
	\item Relation \textit{including} has the highest score 1. The reason is that this relation \textit{(outfit, including, staff)} is the necessary bridge to acquire what staffs an outfit consists of. And the same reason has been reflected in the impact of model depth.
\end{itemize}
 
\section{related work}
Our work is highly related with the knowledge-aware recommendation, which can be grouped into three categories.

\hspace*{\fill} \ 

\noindent\textbf{Embedding-based Methods} \cite{ai2018learning,cao2019unifying,zhang2016collaborative,huang2018improving,zhang2018learning,wang2018dkn} hire KG embedding algorithm (i.e., TransE \cite{bordes2013translating} and TransH \cite{wang2016text}) to model prior representations of item, which are used to guide the recommender model. For example, DKN \cite{wang2018dkn} learns knowledge-level embedding of entities in news content via TransD \cite{ji2015knowledge} for news recommendation. KSR \cite{huang2018improving} utilizes knowledge base information learned with TransE as attribute-level preference to enhance the sequential recommendation. 

\hspace*{\fill} \ 

\noindent\textbf{Path-based Methods} \cite{yu2013collaborative,yu2013recommendation,yu2014personalized,luo2014hete,shi2015semantic} usually predefines a path scheme (i.e., meta-path) and leverages path-level semantic similarities of entities to refine the representations of users and items. For example, MCRec \cite{hu2018leveraging} learns the explicit representations of meta-paths to depict the interaction context of user-item pairs.
RKGE \cite{sun2018recurrent} mines the path relation between user and item automatically and encodes the entire path using a recurrent network to predict user preference towards this item.

\hspace*{\fill} \ 

\noindent\textbf{GNN-based Methods} \cite{wang2019kgat,wang2019knowledge1,qu2019end,sha2019attentive,zhao2019intentgc,wang2021learning} utilizes the message-passing mechanism in graph to aggregate high-order attribute informations into item representation for enhanced and explainable recommendation. KGAT\cite{wang2019kgat} regards user-item interaction as a new relation added to KG and then employees attentive mechanism to propagate attribute information. IntentGC \cite{zhao2019intentgc} reconstructs user-to-user relationships and item-to-item relationships based on KG and proposes a novel graph convolutional network to aggregate the attribute information from neighbors. KGIN \cite{wang2021learning} learn user interest via an attentive combination of attributes and integrates relational
information from multi-hop paths to refine the representations.

\section{conclusion}
In this paper, we study the attribute information in knowledge graphs intending to improve the recommendation performance. On the item side, the proposed AKGAN can learn more high-quality item representation by remaining the independency of attributes and maintaining the weight of high-order significant attribute nodes. On the user side, AKGAN mines user interests towards attributes and provides a personal recommendation. Extensive experiments demonstrate the effectiveness and explainability of AKGAN.

For future work, we plan to investigate how to model the evolving process of user interests towards attributes based on the long-term behavior sequence. Another direction is to explore whether attribute interaction will benefit recommendation since feature interaction has shown great success in click-through rate prediction.
\begin{acks}
The corresponding author Jianhua Tao thanks the support of National Natural Science Foundation of China.
\end{acks}

\bibliographystyle{ACM-Reference-Format}
\bibliography{ref}

\appendix

\section{APPENDIX}

\subsection{Reproducibility Settings}\label{subsec:embedding dimension}


In KG, the number of edges varies acoording to relation type. We can set $d^m$ as a fixed constant value which is the same as that all latent vectors of different fields share the same dimension in FFM\cite{juan2016field}. However, KGs contain hundreds of relation types, which requires huge computational resources both in memory space and time. Table \ref{tab:Edge number of Alibaba-iFashion dataset} shows an example on the Alibaba-iFashion dataset. Therefore, we design variable dimensions of different attribute spaces. The principle is that more edges belonging to one specific relation type bring in a larger dimension of this attribute space. Formally, we make the dimension increases linearly with the edge number as below:
\begin{equation}
	\label{eq:dimension of attribute space}
	d^{m}=\left\{
		  	\begin{array}{lc}
			d_{min}+\displaystyle{\frac{d_{max}-d_{min}}{c}}|r_m|, & |r_m|  \leq \displaystyle{\frac{c}{d_{max}-d_{min}}}\\
			d_{max}, & |r_m| > \displaystyle{\frac{c}{d_{max}-d_{min}}}
			\end{array}
		\right.
\end{equation}
where $d_{max}$, $d_{min}$, $c$, $c$ are hyperpatameters and $|r_m|$ is the number of edges belonging to relation $r_m$, like $|r_0|=260477$ in Alibaba-iFashion.
 
\begin{table}[h]
	\caption{Edge number of Alibaba-iFashion dataset}
	\label{tab:Edge number of Alibaba-iFashion dataset}
	\begin{tabular}{c|c}
		\hline
		\# $r_1$ - $r_{10}$ & 260,477 469 2518 482 1177 343 429 618 143 865 \\
		\# $r_{11}$ - $r_{20}$ & 2443 1572 155 939 221 934 244 187 162 93 \\
		\# $r_{21}$ - $r_{30}$ & 833 161 129 736 22 61 318 142 435 36 \\
		\# $r_{31}$ - $r_{40}$ & 296 459 19 102 73 23 53 321 63 45 \\
		\# $r_{41}$ - $r_{51}$ & 132 7 10 40 42 15 78 11 11 5 6 \\
		\hline
	\end{tabular}
\end{table}

We list the parameter settings of AKGAN on three datasets in Table \ref{tab:Hyperparameter settings of AKGAN}, where the hyperparameters include the learning rate $l_r$, the coefficient $\lambda$ of $L_2$ regularization, the temperature coefficient $\tau$, and three coefficients about dimension  $d_{max}$, $d_{min}$, $c$. We have released our codes, datasets, model parameters, and training logs at https://github.com/huaizepeng2020/AKGAN.

\begin{table}[h]
	\caption{Hyperparameter settings of AKGAN}
	\label{tab:Hyperparameter settings of AKGAN}
	\small
	\begin{tabular}{l|cccccc}
		\hline & $l_r$ & $\lambda$ & $\tau$ &$d_{max}$& $d_{min}$& $c$ \\
		\hline \hline 
		Amazon-Book & $10^{-4}$ & $10^{-5}$ &0.25&32&4&5000\\
		Last-FM & $10^{-4}$ & $10^{-5}$ &0.5&64&16&5000\\
		Alibaba-iFashion & $10^{-4}$ & $10^{-5}$&0.1&64&4&5000 \\
		\hline
	\end{tabular}
\end{table}

\end{document}